\documentclass[
    aps,
    10pt,
    twocolumn,
    superscriptaddress,
    showpacs,
    preprintnumbers,
    amsmath,
    amssymb,
    physrev,
    nofootinbib
]{revtex4-2}

\usepackage{graphicx}
\graphicspath{{../}{}}
\usepackage{dcolumn}
\usepackage{bm}
\usepackage{amsmath}
\usepackage{amsthm}
\usepackage{amssymb}
\usepackage{mathtools}
\usepackage{xcolor}
\usepackage{hyperref}
\usepackage{cleveref}
\usepackage{comment}
\usepackage{titlesec}
\titlespacing*{\section} {0pt}{3ex}{3ex}
\expandafter\def\expandafter\normalsize\expandafter{%
    \normalsize%
    \setlength\abovedisplayskip{10pt}%
    \setlength\belowdisplayskip{10pt}%
    \setlength\abovedisplayshortskip{-8pt}%
    \setlength\belowdisplayshortskip{2pt}%
}

\newcommand{\E}{\mathbb E}
\newcommand{\dd}{\mathrm d}

\newcommand{\one}{\boldsymbol 1}
\newcommand{\transpose}{\mathsf T}
\newcommand{\order}{\mathcal O}
\newcommand{\Dz}{\mathcal D z}
\DeclareMathOperator{\sech}{sech}

\begin{document}

%\title{Generalization modes in Hopfield-like neural Networks}
\title{Inferring Concepts from Noisy Examples in Hopfield-like Neural Networks}

\author{Marco Benedetti}
\email{marco.benedetti4@unibocconi.it}
\affiliation{Università Bocconi, Milan, Italy}
\author{Victor Dotsenko}
\affiliation{Sorbonne Université, CNRS, Laboratoire de Physique Théorique de la Matière Condensée, Paris, France}
\author{Giulia Fischetti}
\affiliation{Ca' Foscari Università Venezia, ZHAW Switzerland}
\author{Enzo Marinari}
\affiliation{ Università di Roma La Sapienza, Rome, Italy}
\author{Gleb Oshanin}
\affiliation{Sorbonne Université, CNRS, Laboratoire de Physique Théorique de la Matière Condensée, Paris, France}

%\date{\today}

\begin{abstract}
We study a variant of the pseudo-inverse learning rule for Hopfield-like Neural Networks, which allows the network to infer archetypal concepts on the basis of a limited number of examples. 
The key novelty is that the new learning rule compensates for correlations between examples at the population level rather than at the level of the particular training set, reducing overfitting of the dataset and leading to generalization.
The mean-field replica theory for this model reveals how this generalization ability is mediated by a multitude of thermodynamic states, with diverse thermodynamic properties, coexisting with the standard Hopfield ones. They appear and vanish through smooth transitions or discontinuous jumps and, interestingly, show much stronger Replica Symmetry Breaking (RSB) effects than the standard Hopfield model, as captured by our 1RSB analysis. Our results, in excellent agreement with numerical simulations, provide deeper insight into the interplay between memory storage and generalization in attractor neural networks.
\end{abstract}

\maketitle
\section{Introduction}
\label{sec:introduction}

The development of models and tools from theoretical physics has greatly contributed to our understanding of neural computation and memory storage. Among these, attractor neural networks have emerged as particularly important frameworks for understanding how the brain might store and retrieve information~\cite{littleExistencePersistentStates1974, amitSpinglassModelsNeural1985, amitModelingBrainFunction1989,BialekEmergenceofBrains}.
The Hopfield model~\cite{hopfieldNeuralNetworksPhysical1982} represents one of the most influential examples of such networks, consisting of binary neurons connected through symmetric pairwise interactions. In this framework, memory patterns are stored as attracting fixed points of the neural dynamics, enabling associative memory retrieval from partial or corrupted inputs. The synaptic weights are typically determined by the Hebbian learning rule~\cite{hebbOrganizationBehaviorNew1950}, which provides local learning based on correlations between neural activities. Relaxing the restriction to pairwise interactions, dense associative-memory models can achieve substantially larger storage capacities through higher-order interaction terms~\cite{krotovDenseAssociativeMemory2016}, at the cost of a less direct interpretation in terms of local pairwise synaptic couplings.
In the context of Attractor Neural Networks, the ability to generalize can be understood as a synthesis process, where the learning rule extracts useful information about general concepts (Archetypes) on the basis of examples \cite{agliariEmergenceConceptShallow2022}. An Attractor Neural Network is said to be generalizing if it builds attractors more correlated to the archetype than to the individual examples provided for it. The distinction between retaining individual experiences and extracting their shared prototype has appeared in distributed-memory models~\cite{mcclellandDistributedMemory1985}, and dense associative memories~\cite{krotovDenseAssociativeMemory2016}. In this setting, the examples pertaining to the same archetype display strong correlations, and the mutual overlaps between examples act as interference noise, dramatically reducing storage capacity and retrieval quality in the standard Hopfield model~\cite{fontanariStorageCorrelated1990}. This has motivated the development of alternative learning algorithms, most notably the pseudo-inverse learning rule~\cite{personnazInformationStorageRetrieval1985, kanterAssociativeRecallMemory1987}, which can perfectly store any set of linearly independent patterns up to the theoretical limit of one pattern per neuron. 
While the pseudo-inverse rule eliminates interference noise, it suffers from two crucial drawbacks: first, it is fundamentally non-local, requiring global knowledge of all patterns to determine each synaptic weight. This contradicts biological plausibility and makes the learning process computationally expensive, as adding new patterns requires recalculating the entire weight matrix. Second, the pseudo-inverse is too effective in building separate attractors for each example, rather than extracting information about the archetype, thus hindering generalization. 
This exposes a basic tension. Leaving correlations untreated produces interference between related examples, whereas removing them exactly can preserve each example too faithfully. Generalization requires an intermediate operation: systematic interference must be reduced without erasing the common structure that defines a class. This has also been noted in \cite{aquaroRecurrentNeuralNetworks2022, agliariHebbianDreamingSmall2024, serricchioDaydreaming2025}, where a parametric procedure inspired by Hebbian Unlearning~\cite{hopfieldUnlearningHasStabilizing1983, vanhemmenIncreasingEfficiencyNeural1990, benedettiSupervisedPerceptronLearning2022a, benedettiEigenvectorDreaming2024}
was devised, interpolating between Hebbian Learning and pseudo-inverse learning to promote generalization over overfitting in pseudo-inverse-based learning rules.

In this paper, we propose a novel approach that combines accurate pattern storage for hierarchically correlated patterns and strong ability to generalize from a limited number of examples. Unlike in the standard pseudo-inverse rule, in our scheme correlations are dealt with by means of the expected correlation structure of the dataset, rather than the empirical one. Hence, our learning mechanism maintains locality of learning and allows for incremental learning. 
This simple modification has profound consequences, increasing the generalization capabilities of the model, and leading to a rich phase behavior, where multiple solution types coexist, corresponding to different generalization strategies.
The paper is organized as follows:  \cref{sec:model} provides the details of our model; \cref{sec:mft} contains our main results, i.e. the study of the multiple generalizing thermodynamic states of the network, within a Replica Symmetric Mean Field analysis; \cref{sec:results} provides an analytical characterization of the novel generalizing states, mapping their rich phenomenology and stability intervals as functions of the number of examples, temperature, dataset quality, and family size $K$; \cref{sec:simulations} contains the comparison between analytical predictions and simulations, as well as comments on Replica Symmetry Breaking effects.

\section{Related Work}
The biological process of consolidation and unlearning during sleep has inspired the introduction of a time dependence in the coupling definition, shown to act as a regularizer, broadening the retrieval phase across a range of dataset qualities and reducing the interference between correlated examples ~\cite{agliariDreamingNeuralNetworks2019a,agliariRegularizationEarlystoppingDreaming2024}. This mechanism progressively suppresses mixture states that arise as superpositions of different stored patterns. If the dreaming phase is halted at the right moment, only the spurious inter-class mixtures, generated by combining samples from different archetypes, are removed, while the intra-class mixtures, originating from combinations of samples within the same archetype, are preserved. These intra-class mixtures are advantageous because they support generalization: their attractor basins naturally merge into broader minima centered on the archetype~\cite{agliariHebbianDreamingSmall2024}. Recent work has also related the steady states generated by Hebbian unlearning to those produced by nonequilibrium, detailed-balance-breaking dynamics~\cite{beheraConnectionHebbian2025}.
Correlated patterns have also motivated local iterative storage prescriptions~\cite{diederichLearningCorrelated1987} and learning rules built from the principal directions of the pattern-correlation matrix~\cite{coombesUsingFeatures1996}. These approaches aim primarily at faithful storage of correlated memories, whereas the present work asks when deliberately retaining part of their common structure produces archetype attractors.
A complementary line of work has shown that correlations can make latent features themselves retrievable~\cite{negriStorageLearning2023} and can generate attractors for previously unseen examples drawn from the same structured distribution~\cite{kalajRandomFeatures2025}. Correlated-pattern Hopfield networks can also be mapped exactly to restricted Boltzmann machines with Gaussian hidden variables, providing a bridge between associative-memory models and trainable latent-variable architectures~\cite{smartMappingHopfield2021}.

%%%%%%%%%%%%%%%%%%%%%%%%%%%%%%%%%%%%%%%%%%%%%%%%%%%%%%%%
\section{Model Description}
\label{sec:model}
% Discuss here: 
% Pattern correlation structure (ultrametricity, archetypes and examples)
% theoretical pseudo-inverse matrix: empirical pseudo-inverse matrix is too good at shattering the examples, leading to lack of generalization

We consider a fully connected network of $N$ binary neurons $\{\sigma_i = \pm 1\}$ ($i = 1, 2, \ldots, N$) described by the Hamiltonian: 
\begin{equation}
\label{eq:hamiltonian}
H[\sigma, J] \; = \; -\frac{1}{2} \sum_{j \neq i}^{N} J_{ij} \sigma_{i} \sigma_{j}.
\end{equation}
The coupling matrix is a variant of the pseudo-inverse learning rule, namely
\begin{equation}
\label{eq:couplings}
J_{ij} =   \frac{1}{N} \sum_{\mu,\nu}^{P} \left(\overline{C}^{-1}\right)_{\mu \nu} \, \xi^{\mu}_{i} \xi^{\nu}_{j}
\end{equation}
where $\{\xi_i^{\mu}=\pm 1\}$ are $P=\alpha N$ examples and $\overline{C_{\mu\nu}}$ is the expectation over the examples statistics of the correlation matrix $\overline{C_{\mu\nu}} \; = \; \mathbb{E}[\xi_{i}^{\mu} \, \xi_{i}^{\nu} ]$. Notice the difference between this prescription and the pseudo-inverse learning rule\footnote{This difference disappears in the case of a sub-extensive number of archetypes, since for fixed \(P\) and \(N\to\infty\),
\(C_{\mu\nu}^{(N)}:=\frac1N\sum_i\xi_i^\mu\xi_i^\nu
\longrightarrow\bar C_{\mu\nu}\)}, which is based on the empirical correlation of the examples $C_{\mu\nu} \; = 1/N \sum_i \xi_{i}^{\mu} \, \xi_{i}^{\nu}$. While the pseudo-inverse learning rule leverages the detailed knowledge of the random examples presented to the network, our prescription is only aware of their average correlation structure. This has a strong impact on the performance of the model, since the pseudo-inverse learning rule is extremely effective in stabilizing individual examples, and leads to a phenomenology identical to that of the Hopfield model, namely attractors corresponding to the examples, rather than the archetypes. In other words, a form of overfitting. Instead, the population inverse removes only the redundancy predicted by the data-generating process. 
We will see how \cref{eq:couplings} allows the network to create other types of attractors, closer to the archetypes than to the individual examples, enhancing the generalization capabilities of the model. Our learning rule is local in space, i.e. the coupling between neurons $i$ and $j$ only depends on the examples activities on these same sites. It is also incremental in the example number, in the sense that, when learning a successive pattern $\omega$, the contributions of older patterns to $J_{ij}$ change in a way that depends only on the population statistics of $\omega$, and not on its particular realization. 

We focus on a hierarchical structure of correlated examples, based on sampling a number $P/K$ of $N$-dimensional i.i.d. configurations (the archetypes), and then corrupting them with multiplicative noise, to get from each a finite number $K$ of examples. In formulas $\xi_i^{\alpha_1 \alpha_2} = \eta_i^{\alpha_1} \zeta_i^{\alpha_1 \alpha_2}$, where $\eta_i^{\alpha_1}$ represents the archetype and $\zeta_i^{\alpha_1 \alpha_2}$ introduces intra-family variations. Iterating the process generates arbitrarily deep hierarchies, in the spirit of \cite{dotsenkoHierarchicalModelMemory1986, pargaUltrametricOrganizationMemories1986, gutfreundHierarchicallyCorrelated1988}. Correlated patterns generated through deeper combinatorial structures have also been represented by multilayer graphical models, with corresponding modifications of the mean-field equations~\cite{mezardMeanFieldMessagePassing2017}. Here, we present the analysis to a one-level hierarchy, while the general case is discussed in the Appendices. We consider the following statistics for the examples:
\begin{align}
    &P(\eta_{i}^{\alpha_{1}}=x) = \frac{1}{2}\big(\delta(x-1) + \delta(x+1)\big), \label{eq:etaprob}\\
    &P(\zeta_{i}^{\alpha_{1}\alpha_{2}}=x) = \frac{1+\rho}{2}\delta(x-1) + \frac{1-\rho}{2}\delta(x+1), \label{eq:chiprob}
\end{align}
where $\alpha_1 = 1, \ldots, P/K$ labels the families and $\alpha_2 = 1, \ldots, K$ labels patterns within each family. The parameter $\rho$ sets the amount of noise driving examples apart from archetypes. 
It follows that $\overline{C_{\mu\nu}}$ has a block-diagonal structure: within each family, patterns have overlap $q = \rho^2$, while patterns from different families are uncorrelated. This block structure is therefore the only family-level information supplied to the learning rule: it specifies which examples are statistically related and their expected similarity, but not their realization-specific fluctuations.

%%%%%%%%%%%%%%%%%%%%%%%%%%%%%%%%%%%%%%%%%%%%%%%%%%%%%%%%
\section{Replica Symmetric Mean-Field Theory}
\label{sec:mft}
The Replica Symmetric (RS) mean-field theory (MFT) analysis of the model is detailed in \cref{app:free energy computation}. Denoting by $\langle \cdot \rangle$ the thermal expectation for a fixed example set, the order parameters of the model are the overlaps $m_{\alpha_1\alpha_2}:=N^{-1}\mathbb{E}[\sum_i \xi^{\alpha_1\alpha_2}_i \langle \sigma_i \rangle]$, the Edwards-Anderson parameter $Q:=N^{-1}\sum_{i}\langle \sigma_i \rangle^2$ and its conjugate parameter $r$. They are determined by solving saddle point equations for a free energy function, see \cref{eq:general-RS-free-energy}. In addition, we will keep track of the overlap between the network state and the archetypes, $O^{\alpha_1}=N^{-1}\mathbb{E}[\sum_i \eta^{\alpha_1}_i \langle \sigma_i \rangle]$. A high value of either kind of overlap implies that the network dynamics has developed attraction basins close to either an example, or an archetype. In other words, it has created a \emph{memory}. When $O^{\alpha_1}>m_{\alpha_1\alpha_2}$, we will say that the Network is “generalizing”, i.e. it is extracting from the examples information about the archetype, rather than focusing on the individual examples.

\subsection{Ansatz on saddle point solutions}
To characterize the interplay between generalization and memorization in the model, we analyze the MFT solutions where the $m$ variables are non-zero within a single archetype class $\alpha_1$. Therefore, we will suppress the archetype index $\alpha_1$ in what follows. Furthermore, we will assume that $m_2 = m_3 = \cdots = m_K$ while $m_1$ is allowed to have a different value. As we will see, this ansatz captures the essential physics while remaining tractable. The physical interpretation is that our description allows the system to single out a specific example $\alpha_2=1$, while all others are treated in a completely symmetric fashion.
MF equations and details about their solutions are reported in \cref{app:RS_free_energy} and \cref{app:1RSB_free_energy}. Within this ansatz, one finds the standard Hopfield model solutions, where $m_2=0$ and $m_1=0, \; Q=0, \; r=0$ for the paramagnetic solution; $m_1=0, \; Q\not= 0, \; r\not= 0$ for the spin-glass solution;  $m_1\neq 0, \; Q\not= 0, \; r\not= 0$ for the Hopfield retrieval solution.

Alongside these solutions, we find a novel and rich set of solutions, which are “generalizing”, in the sense that $O>m_1,\, m_2$. Namely: 
\begin{itemize}
    \item Fully Symmetric (FS) solution $m_1=m_2,\, Q\not= 0, \; r\not= 0$: the network generalizes without any bias towards any of the examples.
    \item Class Representative (CR) solution $m_1>m_2, \, Q\not= 0, \; r\not= 0$: the network shows a preference towards one of the examples corresponding to the archetype;
    \item Outlier Excluding (OE) solution $m_1<m_2, \, Q\not= 0, \; r\not= 0$: a 'leave-one-out' generalization mechanism, where the network aligns symmetrically and preferentially towards $K-1$ examples, while one is regarded as an outlier;
\end{itemize}
This reveals how the network not only balances between memorizing specific examples and inferring archetypes, as noted for similar learning rules in \cite{aquaroRecurrentNeuralNetworks2022,agliariHebbianDreamingSmall2024}, but also generalizes according to a multitude of strategies, corresponding to states with different thermodynamic properties. 
All generalizing solutions are metastable states of the system, while the lowest free energy states are the same as in the Hopfield model as $\alpha$ and $T$ vary. Still, analogously to the Hopfield model, metastable states can perform neural computations, as they are surrounded by extensive free energy barriers, and escape times to the thermodynamically dominant state are exponential in $N$. Hence, a network prepared near an archetype may relax into the corresponding fixed point and display its characteristic overlap profile.

%%%%%%%%%%%%%%%%%%%%%%%%%%%%%%%%%%%%%%%%%%%%%%%%%%%%%%%%%%%%%%%%%%%%%%%%%%%%%%%%%%%%
\section{Analytical Characterization of the Generalizing States}
\label{sec:results}
\begin{figure}[htb]
   \centering
   \includegraphics[trim={0cm 0cm 0cm 2cm}, clip, width=\linewidth]{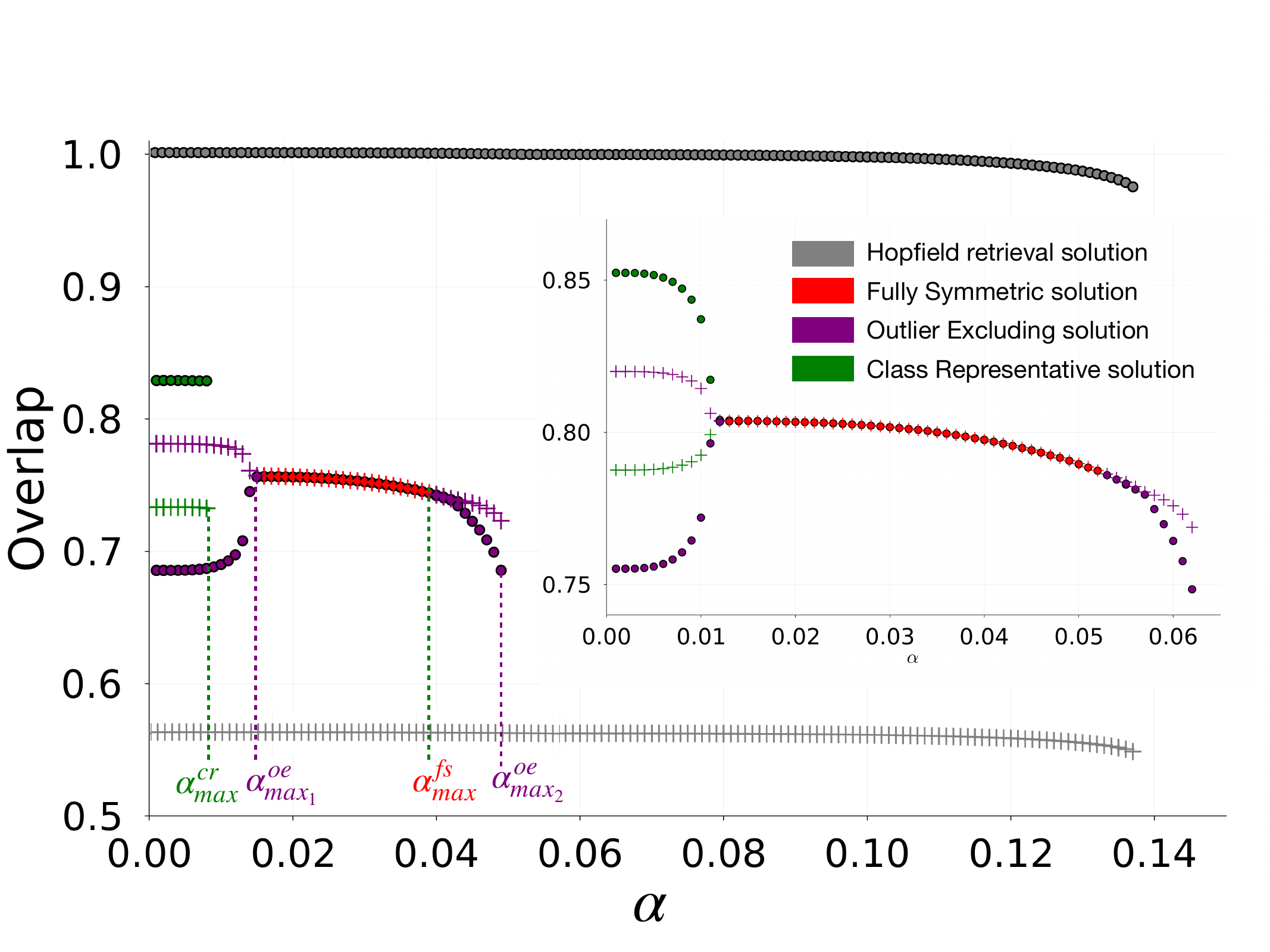}
   \caption{Overlap bifurcation diagrams at $T=0$ and $K=4$. The main figure shows results for $\rho=0.75$, whereas the inset for $\rho=0.8$. The diagrams show different Replica Symmetric solution types characterized by overlaps $m_1$ (dots) and $m_2$ (crosses) as a function of $\alpha$. Red identifies the FS solution, green the CR solution, purple the OE solution. }
\label{fig:bifurcation_comparison}
\end{figure}

\begin{figure}[htb]
   \centering
   \includegraphics[width=\linewidth]{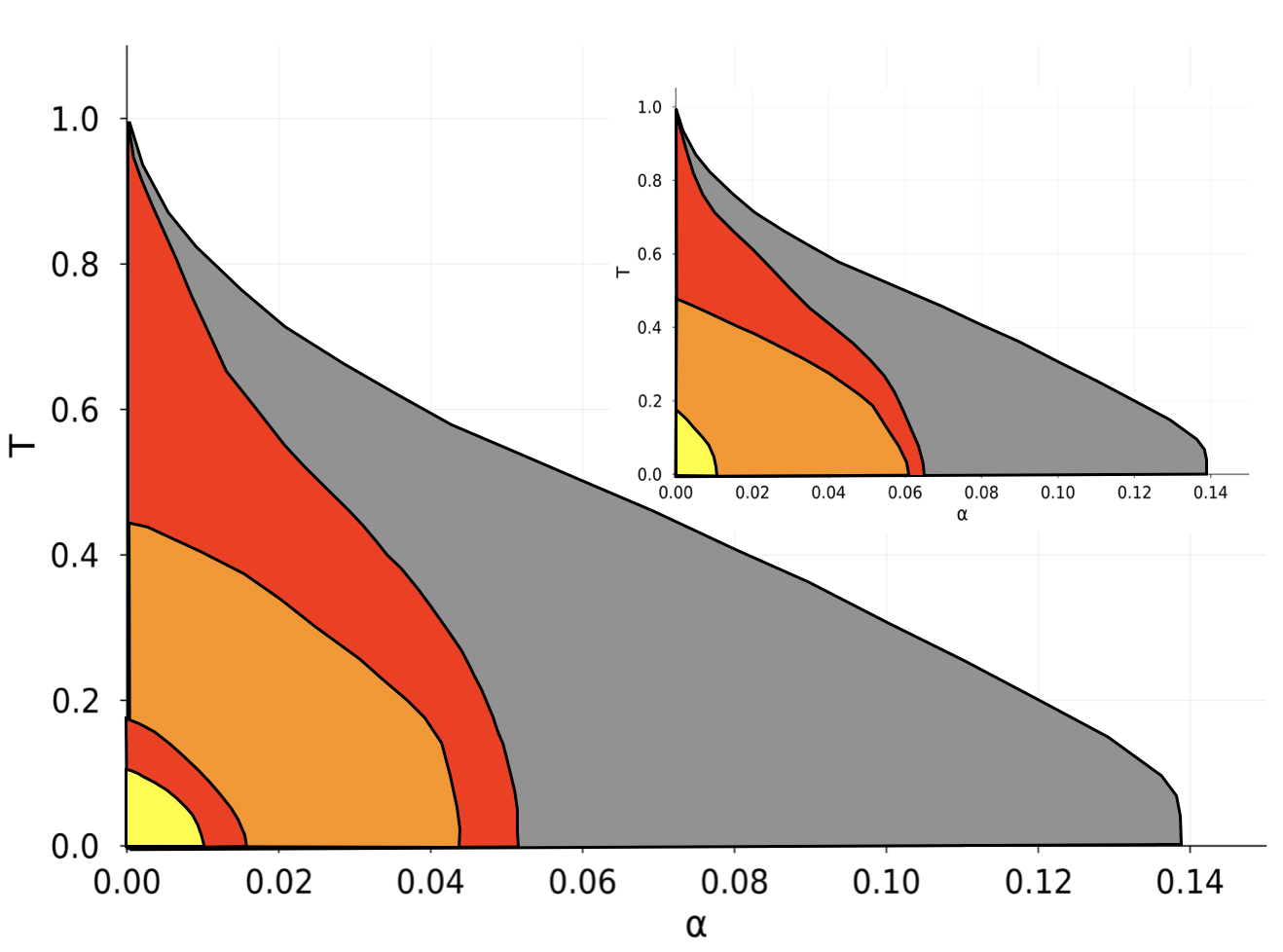}
   \caption{Phase boundaries in the temperature-load parameter space for $K=4$ examples per archetype. Main figure shows results for dataset-quality $\rho=0.75$ while the inset shows results for $\rho=0.80$. Red indicates where only the OE exists, orange indicates where only the FS solution exists, yellow indicates where the CR and OE solution coexists. The gray area indicates the area where only the Hopfield model solutions exists.}
   \label{fig:phase_comparison}
\end{figure}

Having identified the possible organizations of a family, we now ask where each one exists as the load, temperature, data quality, and number of examples are varied. 
The three solutions $m_2=0$, which are also present in the standard Hopfield model, are described by the same phase diagram reported in  \cite{amitSpinglassModelsNeural1985}. Hence, in the following we will only concentrate on the novel $m_2>0$ solutions, where generalization occurs. 
The phenomenology associated with the three kinds of solutions is very rich, as can be appreciated from the zero-temperature picture in the RS approximation in Figure~\ref{fig:bifurcation_comparison}, where the critical loads are color-coded and displayed for ease of reading. It goes as follows: the FS solution exists in an interval $\alpha^{fs}_{min}<\alpha<\alpha^{fs}_{max}$; the CR solution exists in the interval $0<\alpha<\alpha^{cr}_{max}$; the OE solution exists in the intervals $0<\alpha<\alpha^{oe}_{max_1}$ and $\alpha^{oe}_{min}<\alpha<\alpha^{oe}_{max_2}$. In all cases, we found that $0<\alpha^{cr}_{max} < \alpha^{oe}_{max_1}$ and $\alpha^{fs}_{max}=\alpha^{oe}_{min}$.
Varying the values of $T,\,\rho$ and $K$, one can change the values of these thresholds. Broadly speaking, increasing the quality of the dataset leads to a more dominant FS. Specifically, both $\alpha^{cr}_{max}$ and $\alpha^{oe}_{max_1}$ decrease as $\rho$ or K increase, and $\alpha^{oe}_{min}$ increases by increasing $\rho$ or K. Hence the amplitude of the interval $\alpha^{fs}_{min}<\alpha<\alpha^{fs}_{max}$ increases with both $K$and dataset quality, and fills the whole area under the Hopfield transition line when $K\to\infty$ or $\rho\to1^{-}$. The order parameters associated with the FS solution coincide with the Hopfield ones in this limit. Hence, cleaner or more numerous examples favor a symmetric family representation; when the common evidence is weaker, configurations biased toward one example or toward a subset of the family occupy a larger part of the solution diagram.  The solutions also exhibit different behaviors at their critical points: at $\alpha^{oe}_{max_2}$ the OE solution disappears discontinuously. At $\alpha^{oe}_{max_1}$ and $\alpha^{cr}_{max}$ the OE and CR solutions can, depending on $K$and dataset quality $\rho$, either disappear discontinuously, or merge continuously with the FS solution. Figure~\ref{fig:bifurcation_comparison} shows the RS bifurcation diagram at zero-temperature for $K=4$ and $\rho=0.75\,,0.8$. The diagram displays the overlaps $m_1$ (with the singled-out pattern) and $m_2$ (with other patterns in the family) as functions of the load parameter $\alpha$. Different solution types are color-coded, revealing a complex bifurcation structure with multiple solution branches coexisting in certain parameter regions, with smooth transitions between some phases and discontinuous jumps between others.
The temperature-load parameter phase diagrams provide a complete picture of the system's behavior. Figure~\ref{fig:phase_comparison} shows the RS phase boundaries for $K=4$ at two different correlation strengths: $\rho=0.75$ and $\rho=0.8$.

\section{Comparison with Numerical Results and 1RSB}
\label{sec:simulations}
\begin{figure*}[!ht]
   \centering
   \includegraphics[trim={0cm 10cm 0cm 8cm}, clip, width=\textwidth]{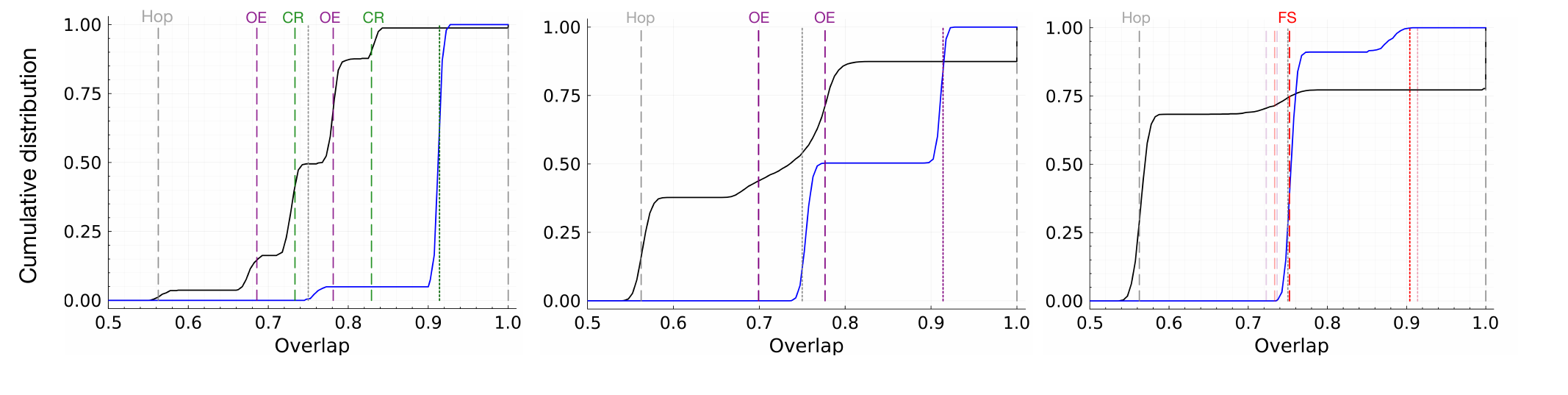}
   \caption{Comparison between 1RSB mean-field theory predictions and simulations for zero-temperature cumulative distribution functions of the overlaps at load parameter $\alpha=0.005$ (left); $\alpha=0.015$ (center); $\alpha=0.045$ (right). Other parameters are: $N=8192$, $K=4$ examples per archetype, correlation strength $\rho=0.75$. Solid lines show cumulative distribution functions of overlaps with examples (black lines) and the archetype (blue lines). MFT predictions are shown as vertical lines, dashed for overlaps with examples, dotted for overlaps with the archetype. Purple corresponds to the OE solution, green to the CR solution, red to the FS solution, gray to the Hopfield solution. In the right panel a light shading indicates the RS estimate, showcasing how 1RSB improves the agreement with simulations.}
   \label{fig:overlap_simulations}
\end{figure*}

We performed numerical simulations to validate our $T=0$ mean-field predictions. We did not compare the finite-temperature picture with simulations, as all generalizing solutions are metastable states of the system, and a thermalized system would always show only the Hopfield solution. 
Simulations are conducted on networks up to $N=8192$ spins, using zero-temperature Monte Carlo dynamics. The network is initialized to one of the archetypes and relaxed to a fixed point, updating all spins sequentially through
\begin{equation}
    \sigma_i^{t+1}=\text{sgn}\Big(\sum_{j\neq i}J_{ij}\sigma_j^{t}\Big)
\end{equation}
We then measure overlaps with both the archetype and all examples within that archetype family, yielding $K+1$ overlaps per simulation (the $K$ example overlaps are collected together). From these overlap distributions, we construct cumulative distribution functions (CDFs) where peaks appear as sharp rises. The step heights reflect relative frequencies and are strongly dependent on initialization strategy, while step locations are physical and comparable to MFT predictions. 
\Cref{fig:overlap_simulations} displays simulation CDFs of overlaps with the examples (black lines) and with the archetypes (blue lines) overlaps at different values of the load parameter $\alpha$. Mean-field theory predictions for the examples and archetype overlaps are shown as dashed and dotted vertical lines, respectively. The color-coding identifies different types of solutions. Notice that different solution types have extremely similar values of archetype overlap, so that the whole zoology of solutions would go undetected if one did not measure the example-overlaps.

Generally speaking, RS MFT gives accurate predictions of the intricate overlap structures displayed by the simulations. While this agreement is essentially perfect for low values of $\alpha$, it degrades and becomes only qualitative as $\alpha$ is increased, particularly close to the transition values of $\alpha$. The 1RSB MFT (see \cref{app:1RSB_free_energy}) predictions are found to always improve the agreement between theory and simulations.
As a rule, 1RSB solutions follow the same qualitative patterns highlighted in \cref{sec:results}, with the difference that all solutions survive to higher values of $\alpha$. This effect, also present in the standard Hopfield model, is more pronounced here, with the zero-temperature critical loads being up to $30\%$  higher than the RS ones. As in the Hopfield model, the RSB becomes relevant only for very low values of $T\sim 0.02$, and leaves the higher-temperature phase diagram unchanged.

\section{Conclusions}
\label{sec:conclusions}
We have analyzed a novel neural network model, using a modified pseudo-inverse learning rule based on theoretical rather than empirical pattern correlations. Replacing empirical covariance with population covariance changes the inductive bias of the pseudo-inverse rule. The network no longer uses the exact geometry of the training set to isolate every example; instead, it can exploit the statistical structure shared by an example family to form collective attractors.
This allows the network to generalize, extracting information about archetypes from a limited number of noisy examples. 
Our RS and 1RSB mean-field analysis reveals a rich solution landscape. We identify six distinct classes of solutions, three of which have generalization properties, meaning that they are characterized by an archetype overlap larger than the overlap with the individual samples. The critical capacities depend on the number of examples per archetype $K$ and the dataset quality $\rho$, with stronger correlations extending the domain of the fully symmetric solution. Interestingly, different solution types exhibit extremely similar archetype overlaps, meaning that the rich solution structure would remain hidden without careful measurement of individual pattern overlaps. The theoretical predictions in the 1RSB approximation are in excellent agreement with zero-temperature Monte Carlo simulations across wide parameter ranges. Replica symmetry breaking effects are more pronounced than in standard Hopfield models, with the RS zero-temperature critical loads being up to $30\%$ lower than the 1RSB ones.

From the point of view of biological plausibility, our model falls into the category of weakly-supervised learning rule, since each example must be correctly assigned to its archetype (and nested sub-categories) to be paired to the correct $(\overline C^{-1})_{\mu\nu}$ indexes. This kind of weak supervision can either be provided by the temporal and contextual organization of experience, where examples for one particular archetype are presented in same time frame or context, or by some simple clustering strategy, where the new memory is categorized based on its empirical correlation with previously learned examples. Moreover, our model assumes prior knowledge of the correlation structure of the dataset. An interesting extension left for future investigation is how would our learning rule perform when new patterns are stored using the empirical correlation of previously learned inputs, removing the need for prior knowledge of the examples second order statistics.
Our work demonstrates that subtle modifications in how neural network learning rules can lead to qualitatively new behaviors, effectively bridging the fundamental tension between memory storage and generalization. This opens promising avenues for both theoretical understanding of neural computation and practical applications in machine learning and computational neuroscience.

\medskip
\bibliography{bibliography}

\onecolumngrid
\appendix
\newpage
\section{free energy computation}
\label{app:free energy computation}
In this section we derive the replica-symmetric mean-field free energy for our model. We consider the case where the dataset comprises a total number of examples $P=\alpha N$, split into $M=P/L$ orthogonal families, and allow for a generic intra-family correlation structure $B$. Thus the complete population covariance has the block form $\overline{C}=I_M\otimes B$.
Let the index $\gamma=1,\ldots,M$ label families, while $\mu=1,\ldots,L$ labels patterns within one family. At site $i$, collect all patterns in family $\gamma$ into $ \boldsymbol\xi_i^\gamma:=\left(\xi_i^{\gamma 1},\ldots,\xi_i^{\gamma L}\right)^{\transpose}\in\{-1,+1\}^{L}$ and denote its common one-site law by $\mu_B$. Then the complete $P\times P$ covariance is
\vspace{0.5cm}
\begin{equation}
    \overline{C_{\gamma\mu,\delta\nu}}=\delta_{\gamma\delta}B_{\mu\nu}, \qquad \overline{C}=I_M\otimes B.
    \label{eq:general-full-covariance}
\end{equation}
i.e.
\begin{align}
    \E_{\boldsymbol\xi_i^\gamma\sim\mu_B}\!\left[\boldsymbol\xi_i^\gamma\right]=\boldsymbol 0, \qquad
    \E_{\boldsymbol\xi_i^\gamma\sim\mu_B}\!\left[\boldsymbol\xi_i^\gamma\boldsymbol\xi_i^{\gamma\transpose}\right]=B
    \label{eq:general-pattern-covariance}\\
\end{align}
The derivation of the free energy follows standard techniques in the theory of disordered systems, and is very similar to the standard Hopfield model (see \cite{amitSpinglassModelsNeural1985}). The partition function of the system reads 
\begin{equation}
    Z = \sum_{\{\sigma\}} \; \exp\biggl\{
\frac{1}{2} \beta \sum_{i,j=1}^{N} J_{ij} \sigma_{i} \sigma_{j} \biggr\} = \sum_{\{\sigma\}=\pm 1} \; \exp\biggl\{
\frac{\beta}{2N} \sum_{i,j=1}^{N} \sum_{\mu,\nu=1}^{P}
\left(\overline{C}^{-1}\right)_{\mu \nu}  \, \xi^{\mu}_{i} \xi^{\nu}_{j} \, \sigma_{i} \sigma_{j} \Big\}
    \label{eq:general-spin-Hamiltonian}
\end{equation}
Performing the normalized Hubbard--Stratonovich transformation gives
\begin{equation}
    Z=\left(\frac{\beta N}{2\pi}\right)^{ML/2}(\det B)^{M/2}
    \int\prod_{\gamma=1}^{M}\dd^L\boldsymbol X_\gamma
    \sum_{\{\sigma\}=\pm1}\exp\left[-\beta H(\boldsymbol X,\boldsymbol\sigma;\boldsymbol\xi)\right].
    \label{eq:general-HS-representation}
\end{equation}
where
\begin{equation}
    H(\boldsymbol X,\boldsymbol\sigma;\boldsymbol\xi):=\frac{N}{2}\sum_{\gamma=1}^{M}\boldsymbol X_\gamma^{\transpose}B\boldsymbol X_\gamma
    -\sum_{\gamma=1}^{M}\sum_{i=1}^{N}\sigma_i\boldsymbol\xi_i^{\gamma\transpose}\boldsymbol X_\gamma.
    \label{eq:general-hidden-Hamiltonian}
\end{equation}
In the spirit of replica calculations, we to compute averages of powers of $Z$, and obtain the quenched free energy through
\begin{equation}
    \E_{\boldsymbol\xi}[\log Z]=\lim_{n\to0}\frac{\E_{\boldsymbol\xi}[Z^n]-1}{n}.
\end{equation}
As in the Hopfield model, assume that only family $\gamma=1$ is condensed. We use the scaling $\boldsymbol X_1^a=\boldsymbol x^a=\order(1), \quad \boldsymbol X_\gamma^a=N^{-1/2}\boldsymbol y_\gamma^a, \quad \gamma=2,\ldots,M.$
Independence and identical distribution of the families then give
\begin{align}
    \E_{\boldsymbol\xi}Z^n &\propto \int\prod_{a=1}^{n}\dd^L\boldsymbol x^a
    \prod_{i=1}^{N}\prod_{a=1}^{n}\left(\sum_{\sigma_i^a=\pm1}\right)
    \exp\left(-\frac{\beta N}{2}\sum_{a=1}^{n}\boldsymbol x^{a\transpose}B\boldsymbol x^a\right) \prod_{i=1}^{N}\E_{\boldsymbol\xi_i^1}
    \exp\left(\beta\sum_{a=1}^{n}\sigma_i^a\boldsymbol\xi_i^{1\transpose}\boldsymbol x^a\right)
    \left[\widetilde Z_n[\boldsymbol\sigma;B]\right]^{M-1}.
    \label{eq:general-replicated-separated-families}
\end{align}
where the contribution of one representative non-condensed family is
\begin{align}
    \widetilde Z_n[\boldsymbol\sigma;B] &:= \left(\frac{\beta}{2\pi}\right)^{nL/2}(\det B)^{n/2}
    \int\prod_{a=1}^{n}\dd^L\boldsymbol y^a
    \exp\left(-\frac{\beta}{2}\sum_{a=1}^{n}\boldsymbol y^{a\transpose}B\boldsymbol y^a\right)\prod_{i=1}^{N}\E_{\boldsymbol\xi}
    \exp\left(\frac{\beta}{\sqrt N}\sum_{a=1}^{n}\sigma_i^a\boldsymbol\xi_i^{\transpose}\boldsymbol y^a\right).
    \label{eq:general-one-family-before-average}
\end{align}
The $\widetilde Z_n[\boldsymbol\sigma;B]$ term can be computed by expanding
\begin{align}
    &\E_{\boldsymbol\xi}
    \exp\left(\frac{\beta}{\sqrt N}\sum_a\sigma_i^a\boldsymbol\xi^{\transpose}\boldsymbol y^a\right)=1+\frac{\beta^2}{2N}\sum_{a,c=1}^{n}\sigma_i^a\sigma_i^c
    \boldsymbol y^{a\transpose}B\boldsymbol y^c+\order(N^{-2}).
    \label{eq:general-single-site-cumulant}
\end{align}
The $N^{-1/2}$ term vanishes because of the zero mean, and all odd orders vanish by central symmetry. Using $\prod_i(1+A_i/N)=\exp[N^{-1}\sum_iA_i+\order(N^{-1})]$, we obtain 
\begin{align}
    \widetilde Z_n[\boldsymbol\sigma;B] &=\left(\frac{\beta}{2\pi}\right)^{nL/2}(\det B)^{n/2}
    \int\prod_{a=1}^{n}\dd^L\boldsymbol y^a
    \exp\left[-\frac{1}{2}\sum_{a,c=1}^{n}\mathcal A_{ac}\boldsymbol y^{a\transpose}B\boldsymbol y^c+\order(N^{-1})\right].
    \label{eq:general-one-family-Gaussian}
\end{align}
where 
\begin{equation}
    Q_{ac}:=\frac{1}{N}\sum_{i=1}^{N}\sigma_i^a\sigma_i^c, \qquad Q_{aa}=1, \qquad \mathcal A_{ac}:=\beta\delta_{ac}-\beta^2Q_{ac}. \label{eq:general-replica-overlap}
\end{equation}
To evaluate this quantity, stack the $n$ vectors $\boldsymbol y^a$ into a single $nL$-dimensional vector. The quadratic form in \cref{eq:general-one-family-Gaussian} then has precision matrix $\mathcal A\otimes B$, whose determinant is
\begin{equation}
    \det(\mathcal A\otimes B)=(\det\mathcal A)^L(\det B)^n.
    \label{eq:general-Kronecker-determinant}
\end{equation}
Applying the standard $nL$-dimensional Gaussian integral therefore gives
\begin{align}
    \widetilde Z_n[\mathsf Q;B] &=\left(\frac{\beta}{2\pi}\right)^{nL/2}(\det B)^{n/2}(2\pi)^{nL/2}
    \left[\det(\mathcal A\otimes B)\right]^{-1/2} =\beta^{nL/2}(\det\mathcal A)^{-L/2}.
    \label{eq:general-one-family-determinant}
\end{align}
The factors $2\pi$ and $\det B$ cancel exactly, which is why the extensive non-condensed contribution has no remaining explicit dependence on the intra-family matrix $B$. Since there are $M-1=\alpha N/L+\order(1)$ non-condensed families,
\begin{equation}
    \left[\widetilde Z_n[\mathsf Q;B]\right]^{M-1}=\exp\left[\frac{\alpha N}{L}\log\widetilde Z_n[\mathsf Q;B]+o(N)\right].
    \label{eq:general-all-noncondensed-families}
\end{equation}

We now enforce the definition \cref{eq:general-replica-overlap} by inserting the integral representation of the $\delta$ function into \cref{eq:general-replicated-separated-families} 
\begin{align}
    1&\propto\int\prod_{a<c}\dd Q_{ac}\,\dd s_{ac}
    \exp\left[2iN\sum_{a<c}s_{ac}Q_{ac}
    -2i\sum_{a<c}s_{ac}\sum_{i=1}^{N}\sigma_i^a\sigma_i^c\right]
    \nonumber\\ &=\int\prod_{a<c}\dd Q_{ac}\,\dd s_{ac}
    \exp\left[iN\sum_{a\neq c}s_{ac}Q_{ac}
    -i\sum_{a\neq c}s_{ac}\sum_{i=1}^{N}\sigma_i^a\sigma_i^c\right].
    \label{eq:general-Q-constraint}
\end{align}
and average over the disorder, obtaining
\begin{align}
    \E_{\boldsymbol\xi}Z^n &\propto\int\dd\mathsf Q\,\dd\mathsf s\prod_{a=1}^{n}\dd^L\boldsymbol x^a
    \exp\left[N\Phi_n(\mathsf Q,\mathsf s,\{\boldsymbol x^a\})+o(N)\right],
    \label{eq:general-replicated-partition-function}\\
    \Phi_n &:=-\frac{\beta}{2}\sum_{a=1}^{n}\boldsymbol x^{a\transpose}B\boldsymbol x^a
    +i\sum_{a\neq c}s_{ac}Q_{ac}
    +\frac{\alpha}{L}\log\widetilde Z_n[\mathsf Q;B]
    +\log\mathcal Z_1(\mathsf s,\{\boldsymbol x^a\}).
    \label{eq:general-replicated-action}
\end{align}
where the condensed-family single-site factor is
\begin{align}
    \mathcal Z_1(\mathsf s,\{\boldsymbol x^a\}) &:=\sum_{\sigma^1,\ldots,\sigma^n=\pm1}\E_{\boldsymbol\xi}
    \exp\left[\beta\sum_{a=1}^{n}\sigma^a\boldsymbol\xi^{\transpose}\boldsymbol x^a
    -i\sum_{a\neq c}s_{ac}\sigma^a\sigma^c\right].
    \label{eq:general-condensed-site-factor}
\end{align} 

\newpage
\section{Replica-symmetric approximation}
\label{app:RS_free_energy}
The replica-symmetric-approximation states that the saddle-point values of the order parameter matrices $Q_{ab}$ and $s_{ab}$, as well as the components of the replica vector $x^{a}_{\alpha}$, do not depend on the replica indices. The saddle-point value of $s$ is purely imaginary, so we introduce the real parameter $r$ through the ansatz
\begin{equation}
    Q_{a\neq b}=Q, \qquad s_{a\neq b}=s=\frac{i}{2}\alpha\beta^2r, \qquad x^a_\alpha=x_\alpha.
    \label{eq:general-RS-ansatz}
\end{equation}
The replica-overlap matrix and the replica-space precision become
\begin{align}
    \mathsf Q &=(1-Q)I_n+QJ_n, \qquad J_n:=\one_n\one_n^{\transpose},
    \label{eq:general-RS-Q-matrix}\\
    \mathcal A &=\beta I_n-\beta^2\mathsf Q=\beta D I_n-\beta^2QJ_n,
    \label{eq:general-RS-precision}\\
    D &:=1-\beta(1-Q).
    \label{eq:general-D-definition}
\end{align}
 $J_n$ has one eigenvalue $n$ and $n-1$ zero eigenvalues, leading to
\begin{align}
    \det\mathcal A &=(\beta D)^{n-1}\left(\beta D-n\beta^2Q\right)
    \nonumber\\ &=(\beta D)^n\left(1-\frac{n\beta Q}{D}\right).
    \label{eq:general-det-RS-precision}
\end{align}
and
\begin{equation}
    \lim_{n\to0}\frac{1}{n}\log\widetilde Z_n[\mathsf Q;B]=-\frac{L}{2}\left[\log D-\frac{\beta Q}{D}\right].
    \label{eq:general-one-family-replica-limit}
\end{equation}
Under \cref{eq:general-RS-ansatz}, we have
\begin{align}
    i\sum_{a\neq c}s_{ac}Q_{ac}&=\frac{1}{2}\alpha\beta^2rQ\,n+\order(n^2),
    \label{eq:general-RS-conjugate-term}\\
    -i\sum_{a\neq c}s_{ac}\sigma^a\sigma^c&=\frac{1}{2}\alpha\beta^2r\left[\left(\sum_{a=1}^{n}\sigma^a\right)^2-n\right].
    \label{eq:general-RS-spin-coupling}
\end{align}
We decouple the square using
\begin{equation}
    \exp\left[\frac{1}{2}\alpha\beta^2r\left(\sum_a\sigma^a\right)^2\right]=\int\Dz\,\exp\left(\beta\sqrt{\alpha r}\,z\sum_a\sigma^a\right),
    \label{eq:general-RS-spin-decoupling}
\end{equation}
and obtain
\begin{align}
    \mathcal Z_1^{\mathrm{RS}}&=e^{-n\alpha\beta^2r/2}\int\Dz\,\E_{\boldsymbol\xi}
    \left[2\cosh\left(\beta h(\boldsymbol\xi,z;\boldsymbol x,r)\right)\right]^n.
    \label{eq:general-RS-site-factor}\\
    \lim_{n\to0}\frac{1}{n}\log\mathcal Z_1^{\mathrm{RS}} &=-\frac{1}{2}\alpha\beta^2r +\int\Dz\,\E_{\boldsymbol\xi}\log\left[2\cosh\left(\beta h(\boldsymbol\xi,z;\boldsymbol x,r)\right)\right].
    \label{eq:general-RS-site-limit}
\end{align}
where the effective field is $h(\boldsymbol\xi,z;\boldsymbol x,r):=\boldsymbol\xi^{\transpose}\boldsymbol x+\sqrt{\alpha r}\,z$. Combining \cref{eq:general-replicated-action,eq:general-one-family-replica-limit,eq:general-RS-conjugate-term,eq:general-RS-site-limit}, we obtain the free energy of the diagonal-including Hamiltonian~\eqref{eq:general-spin-Hamiltonian}:
\begin{align}
    f_{\mathrm{RS}}[\boldsymbol x,Q,r] &=\frac{1}{2}\boldsymbol x^{\transpose}B\boldsymbol x
    +\frac{1}{2}\alpha\beta r(1-Q)
    +\frac{\alpha}{2\beta}\left[\log D-\frac{\beta Q}{D}\right]
    \nonumber\\
    &\quad-\frac{1}{\beta}\int\Dz\,\E_{\boldsymbol\xi}
    \log\left[2\cosh\left(\beta h(\boldsymbol\xi,z;\boldsymbol x,r)\right)\right].
    \label{eq:general-RS-free-energy}
\end{align}

Stationarity of \cref{eq:general-RS-free-energy} with respect to $x_\mu$ gives the vector equation
\begin{equation}
    B\boldsymbol x=\int\Dz\,\E_{\boldsymbol\xi}\left[
    \boldsymbol\xi\tanh\left(\beta h(\boldsymbol\xi,z;\boldsymbol x,r)\right)\right].
    \label{eq:general-x-saddle}
\end{equation}
Stationarity with respect to $Q$ gives
\begin{equation}
    r=\frac{Q}{D^2}=\frac{Q}{\left[1-\beta(1-Q)\right]^2}.
    \label{eq:general-r-saddle}
\end{equation}
Stationarity with respect to $r$ gives
\begin{equation}
    0=\frac{\partial f_{\mathrm{RS}}}{\partial r}=\frac{\alpha\beta}{2}(1-Q)-\frac{\sqrt\alpha}{2\sqrt r}
    \int\Dz\,\E_{\boldsymbol\xi}\left[z\tanh\left(\beta h\right)\right].
    \label{eq:general-r-derivative}
\end{equation}
Gaussian integration by parts yields
\begin{equation}
    \int\Dz\,\E_{\boldsymbol\xi}\left[z\tanh(\beta h)\right]=\beta\sqrt{\alpha r}\int\Dz\,\E_{\boldsymbol\xi}\left[\sech^2(\beta h)\right].
    \label{eq:general-Gaussian-IBP}
\end{equation}
Substitution in \cref{eq:general-r-derivative} gives
\begin{align}
    1-Q &=\int\Dz\,\E_{\boldsymbol\xi}\left[\sech^2\left(\beta h(\boldsymbol\xi,z;\boldsymbol x,r)\right)\right],
    \label{eq:general-Q-saddle-sech}\\
    Q &=\int\Dz\,\E_{\boldsymbol\xi}\left[\tanh^2\left(\beta h(\boldsymbol\xi,z;\boldsymbol x,r)\right)\right].
    \label{eq:general-Q-saddle-tanh}
\end{align}
Together, \cref{eq:general-x-saddle,eq:general-r-saddle,eq:general-Q-saddle-tanh} form the complete finite-temperature RS saddle-point system before imposing any symmetry on $\boldsymbol x$.

\section{Interpretation of the order parameters}
The preceding calculation introduced \(\boldsymbol x\), \(Q\), and \(r\) as saddle-point variables. We now reconnect them to physical observables. Define the physical overlap with the patterns in the condensed family by
\begin{equation}
    \boldsymbol m:=\lim_{N\to\infty}\frac{1}{N}\sum_{i=1}^{N}
    \E_{\boldsymbol\xi}\left[\boldsymbol\xi_i^1\left\langle\sigma_i\right\rangle_{\boldsymbol\xi}\right].
    \label{eq:general-pattern-overlap-vector}
\end{equation}
The relation between $m$ and $X$ is understood through the identity
\begin{equation}
    \boldsymbol0=\left\langle\frac{\partial H}{\partial\boldsymbol X_1}\right\rangle_{\boldsymbol\xi}=NB\left\langle\boldsymbol X_1\right\rangle_{\boldsymbol\xi}
    -\sum_{i=1}^{N}\boldsymbol\xi_i^1\left\langle\sigma_i\right\rangle_{\boldsymbol\xi}.
    \label{eq:general-condensed-X-integration-by-parts}
\end{equation}
The first equality comes by the fundamental theorem of calculus applied to the vector field $e^{-\beta H(\boldsymbol X,\boldsymbol\sigma;\boldsymbol\xi)}\,\boldsymbol e_{1\mu}$
where \(\boldsymbol e_{1\mu}\) is a versor pointing in the \(X_{1\mu}\) direction, and the second equality comes from \cref{eq:general-HS-representation}. After division by $N$, the thermodynamic limit gives the exact relation $\boldsymbol m=B\boldsymbol x.$ To understand the meaning of the $r$ parameter, consider the extremality wrt $Q_{ac}$ in \cref{eq:general-replicated-action}
\begin{equation}
    0=\frac{\partial\Phi_n}{\partial Q_{ac}}=2is_{ac}+\frac{\alpha\beta^2}{L}
    \left\langle\boldsymbol y^{a\transpose}B\boldsymbol y^c\right\rangle_{\widetilde Z_n}.
    \label{eq:general-Qac-variation}
\end{equation}
Substitution of $s_{a\neq c}=i\alpha\beta^2r/2$ gives, for $a\neq c$,
\begin{equation}
    r=\frac{1}{L}\left\langle\boldsymbol y^{a\transpose}B\boldsymbol y^c\right\rangle_{\widetilde Z_n},
    \label{eq:general-r-hidden-overlap}
\end{equation}
where $\langle\cdot\rangle_{\widetilde Z_n}$ is the normalized average defined by the Gaussian measure in \cref{eq:general-one-family-Gaussian}. For any representative non-condensed family $\gamma>1$, recall that $\boldsymbol y^a=\sqrt N\,\boldsymbol X_\gamma^a$. Therefore,
\begin{align}
    \left\langle\boldsymbol X_\gamma^{a\transpose}B\boldsymbol X_\gamma^c\right\rangle_{\boldsymbol\xi} &=\left\langle\boldsymbol X_\gamma^a\right\rangle_{\boldsymbol\xi}^{\transpose}B
    \left\langle\boldsymbol X_\gamma^c\right\rangle_{\boldsymbol\xi},\quad a\neq c
    \label{eq:general-replica-factorization-X}
\end{align}
and \cref{eq:general-r-hidden-overlap} becomes
\begin{equation}
    r=\lim_{N\to\infty}\frac{N}{L}\E_{\boldsymbol\xi}\left[
    \left\langle\boldsymbol X_\gamma\right\rangle_{\boldsymbol\xi}^{\transpose}B
    \left\langle\boldsymbol X_\gamma\right\rangle_{\boldsymbol\xi}\right].
    \label{eq:general-r-representative-family}
\end{equation}
Since $M-1=\alpha N/L+\order(1)$ and the non-condensed families are exchangeable, this is equivalently
\begin{equation}
    r=\lim_{N\to\infty}\frac{1}{\alpha}\sum_{\gamma=2}^{M}
    \E_{\boldsymbol\xi}\left[
    \left\langle\boldsymbol X_\gamma\right\rangle_{\boldsymbol\xi}^{\transpose}B
    \left\langle\boldsymbol X_\gamma\right\rangle_{\boldsymbol\xi}\right].
    \label{eq:general-r-all-hidden-families}
\end{equation}
or equivalently
\begin{equation}
    r=\lim_{N\to\infty}\frac{1}{\alpha}\sum_{\gamma=2}^{M}
    \E_{\boldsymbol\xi}\left[
    \widehat{\boldsymbol m}_\gamma^{\transpose}B^{-1}\widehat{\boldsymbol m}_\gamma\right].
    \label{eq:general-r-pattern-overlaps}
\end{equation}
This shows that $r$ measures the total squared overlap with the non-condensed families in the metric $B^{-1}$.

\section{Specialization to a one-depth hierarchy}
In this work, the MFT predictions are compared with simulations for a depth-one hierarchy of examples. In formulas $\xi_i^{\alpha_1 \alpha_2} = \eta_i^{\alpha_1} \zeta_i^{\alpha_1 \alpha_2}$, where $\eta_i^{\alpha_1}$ represents the archetype and $\zeta_i^{\alpha_1 \alpha_2}$ introduces intra-family variations:
\begin{align}
    &P(\eta_{i}^{\alpha_{1}}=x) = \frac{1}{2}\big(\delta(x-1) + \delta(x+1)\big), \label{eq:appendix-etaprob}\\
    &P(\zeta_{i}^{\alpha_{1}\alpha_{2}}=x) = \frac{1+\rho}{2}\delta(x-1) + \frac{1-\rho}{2}\delta(x+1). \label{eq:appendix-chiprob}
\end{align}
Here $\alpha_1 = 1, \ldots, P/K$ labels the families and $\alpha_2 = 1, \ldots, K$ labels patterns within each family. The parameter $\rho$ sets the amount of noise driving examples apart from archetypes. 
This corresponds to $\overline{C_{\mu\nu}}$ having a block-diagonal structure: within each family, patterns have overlap $q = \rho^2$, while patterns from different families are uncorrelated. SpecifySpecializinging the results of the above sections to this case, we have
\begin{equation}
\langle Z^n\rangle \, \propto \, 
\exp\Bigl\{-\beta \, n \, N \; F[Q, \, r, \, x_{1}, \ldots, x_{K} ] \Bigr\}
\label{A10}
\end{equation}
with
\begin{equation}
\begin{split}
F[Q, \, r, \, x_{1}, \ldots, x_{K}] &=
\frac{1}{2} (1-q) 
\sum_{\alpha=1}^{K} \bigl(x_{\alpha}\bigr)^{2} 
+ \frac{1}{2} q \, \biggl(\sum_{\alpha=1}^{K} x_{\alpha}\biggr)^{2}
+ \frac{1}{2} \alpha\beta \, r \, (1- Q) 
\\
&-
\frac{1}{\beta}
\int {\cal D} z \, \Bigl\langle\Bigl\langle
\ln\biggl[
2\cosh\Bigl[ 
\beta\Bigl(\sum_{\alpha=1}^{K} \zeta^{\alpha} x_{\alpha}
+ \sqrt{\alpha r} \, z \, \Bigr)
\Bigr]
\biggr]
\Bigr\rangle\Bigr\rangle_{ \zeta}
\\
&+
\frac{\alpha}{2\beta}
\biggl[
\ln\bigl[1 - \beta(1-Q)\bigr]
-\frac{\beta \, Q}{1 - \beta(1-Q) }
\biggr].
\end{split}
\label{eq:RS_F_T!=0}
\end{equation}
In the second line, the disorder is a random walk, with the "steps" $\zeta$ distributed according to $P(\zeta^{\alpha}=x) = \frac{1+\rho}{2}\delta(x-1) + \frac{1-\rho}{2}\delta(x+1)$, and the average computation is detailed in \cref{APP:Average_over_disorder}. Taking the derivatives with respect to $x_\alpha$, $Q$ and $r$ we obtain the corresponding saddle point equations:
\begin{equation}
    \label{eq:RS_finiteT_speq}
    \begin{aligned}
        &(1-q) \, x_{\alpha} \, + \, q \, \Bigl( \sum_{\alpha'=1}^{K} x_{\alpha'}\Bigr)
        \, (=m^\alpha) = \, \int {\cal D} z \bigl\langle\bigl\langle \,\zeta^{\alpha}
        \tanh\Bigl(\beta\Bigl[\Big(\sum_{\alpha=1}^{K} \zeta^\alpha x_{\alpha}\Big) +\sqrt{\alpha r} \, z\Big]\Bigr) \bigr\rangle\bigr\rangle_{ \zeta}\\
        &r \; = \;   \frac{Q}{\Bigl(1 - \beta(1-Q)\Bigr)^{2}} \\
        &(1-Q) \; = \; \int {\cal D} z \, \bigl\langle\bigl\langle \cosh^{-2} \Bigl(\beta\Bigl[\Big(\sum_{\alpha=1}^{K} \zeta^\alpha x_{\alpha}\Big) +\sqrt{\alpha r}\, z\Big]\Bigr) \bigr\rangle\bigr\rangle_{\zeta}
        \end{aligned}
\end{equation}
The main text reports the analysis of the solutions to this set of equations, under the additional symmetry assumption $x_2=x_3=\ldots=x_K$, while $x_1$ is free. Once the values of $x_1$ and $x_2$ are found, the archetype overlaps in the RS approximation are given by
\begin{equation}
    \label{eq:archoverl}
    O=\int {\cal D} z \bigl\langle\bigl\langle \, \eta
\tanh\Bigl(\beta\Bigl[\Big(\sum_{\alpha=1}^{K} \xi^\alpha x_{\alpha}\Big) +\sqrt{\alpha r} \, z\Big]\Bigr) \bigr\rangle\bigr\rangle_{\eta,\xi^1,\ldots,\xi^K}
\end{equation}
\section{Zero Temperature Replica Symmetric free energy}
In the $T\to0$ limit, one can easily see that:
\begin{equation}
    \begin{aligned}
        &\lim_{\beta\to\infty}  \biggl\{\tanh\Bigl(\beta t \Bigr) \biggr\}\; = \; \mbox{sign} \Bigl(t\Bigr) \\
        &\lim_{\beta\to\infty} \biggl\{\frac{\beta}{2} \, \cosh^{-2} \Bigl(\beta t\Bigr)\biggr\}\; = \; \delta\Bigl(t\Bigr)
    \end{aligned}
\end{equation}
Then, introducing the notation $\beta(1-Q) = C$ and noting that in the
limit $\beta\to\infty$ $\; \; Q \to 1$, from \cref{eq:RS_finiteT_speq}
we get:
\begin{equation}
    \begin{aligned}
        &(1-q) \, x_{\alpha} \, + \, q \, \Bigl(x_{1} + \sum_{\alpha'=2}^{K} x_{\alpha'}\Bigr) = \Bigl\langle\Bigl\langle \, \zeta^{\alpha} \mbox{erf} \biggl(\frac{1}{\sqrt{2\alpha r}}\Bigl[\sum_{\alpha'=1}^{K} \zeta^{\alpha'}x_{\alpha'}\Bigr]\biggr)\Bigr\rangle\Bigr\rangle_{\zeta} \\
        &r = \frac{1}{ \bigl(1 - C\bigr)^{2}} \\
        &C = \sqrt{\frac{2}{\pi\alpha r}} \, \Bigl\langle\Bigl\langle\exp\biggl\{-\frac{1}{2\alpha r}\Bigl[\sum_{\alpha'=1}^{K} \zeta^{\alpha'} x_{\alpha'}\Bigr]^{2}\biggr\}\Bigr\rangle\Bigr\rangle_{\zeta}
    \end{aligned}
    \label{eq:RS_zeroT}
\end{equation}
where $\mbox{erf}(x)$ is the error function $\mbox{erf}(x) \; = \; \frac{2}{\sqrt{\pi}}\int_{0}^{x} dz \, \exp\bigl(- z^{2}\bigr)$.
\section{Finite Temperature 1RSB free energy}
\label{app:1RSB_free_energy}
As for the case of the RS free energy, the computation of the 1RSB approximation of our model follows very closely the standard Hopfield-model computation, see for example \cite{crisantiSaturationLevelHopfield1986a, steffanReplicaSymmetryBreaking1994}. Starting from \cref{eq:general-replicated-action,eq:general-one-family-determinant,eq:general-condensed-site-factor}, we now consider the 1RSB structure of the overlap matrix 
\begin{equation}
    \begin{aligned}
        &\boldsymbol{Q}=(1-q_1)\mathbb{I}_n + (q_1-q_0)\mathbb{I}_\frac{n}{m}\otimes\boldsymbol{e}_m\boldsymbol{e}_m^T + q_0\boldsymbol{e}_n\boldsymbol{e}_n^T\\
        &\boldsymbol{s}=(1-s_1)\mathbb{I}_n + (s_1-s_0)\mathbb{I}_\frac{n}{m}\otimes\boldsymbol{e}_m\boldsymbol{e}_m^T + s_0\boldsymbol{e}_n\boldsymbol{e}_n^T,
    \end{aligned}
\end{equation}
where $\mathbb{I}_n$ denotes an $n$-dimensional identity matrix and $\boldsymbol{e}_m$ is an $m$-dimensional column vectors of ones, so that $\boldsymbol{e}_m\boldsymbol{e}_m^T$ is a $m\times m$ matrix of ones. In the limit $n\to 0$, the variable $m$ gets promoted to a real variable $0\leq m\leq1$. After changing variables $s \; = \; \frac{i}{2} \alpha \beta^{2} \, r$, we obtain (omitting irrelevant constants):
\begin{align}
&\lim_{n\to 0} 
\frac{1}{n} \ln\bigl[\tilde{{\cal Z}}_{n, K} \bigl(r; \{x_{\alpha}\})\bigr]
\; = \; \frac{1}{m} 
\int {\cal D} z \Big\langle\Big\langle\log \int {\cal D} \tilde{z} \Big[2\cosh \Big( \beta\big(\sum_{\alpha=1}^{K} \zeta^{\alpha} x_{\alpha}
+ \sqrt{\alpha r_0} \, z + \sqrt{\alpha(r_1-r_0)}\tilde{z}\, \big)  \Big)   \Big]^m\Big\rangle\Big\rangle_\zeta\\
&\lim_{n\to 0} \, \frac{1}{n} \ln\bigl[\tilde{I}_{n, K} \bigl(Q)\bigr] = -\frac{1}{2} K \biggl[
\left(1-\frac{1}{m}\right)\log(1-\beta+\beta q_1)+ \frac{1}{m}\log\Big(1-\beta+\beta\left(mq_0+(1-m)q_1\right) \Big) \\ \nonumber 
&\qquad\qquad\qquad\qquad\qquad-\frac{\beta q_0}{1-\beta+\beta\big(mq_0+(1-m)q_1\big)}\Big]
\end{align} 
Finally, the expression for the free energy as a function of the order parameters reads
\begin{equation}
\label{eq:1RSB_F_T!=0}
\begin{split}
F[q_0,q_1,r_0,r_1,m, \, x_{1}, \ldots, x_{K}] &=
\frac{1}{2} (1-q) 
\sum_{\alpha=1}^{K} \bigl(x_{\alpha}\bigr)^{2} 
+ \frac{1}{2} q \, \biggl(\sum_{\alpha=1}^{K} x_{\alpha}\biggr)^{2}
- \frac{1}{2}\alpha\beta \Big(r_1q_1(1-m)+r_0q_0m-r_1\Big)
\\
&-
\frac{1}{\beta m}
\int {\cal D} z \Big\langle\Big\langle \log \int {\cal D} \tilde{z} \Big[2\cosh \Big(\beta(\sum_{\alpha=1}^{K} \zeta^{\alpha} x_{\alpha}
+ \sqrt{\alpha r_0} \, z + \sqrt{\alpha(r_1-r_0)}\tilde{z}\, \big)  \Big)   \Big]^m\Big\rangle\Big\rangle_{\zeta}
\\
&+
\frac{\alpha}{2\beta}
\biggl[
\left(1-\frac{1}{m}\right)\log(1-\beta+\beta q_1)+ \frac{1}{m}\log\Big(1-\beta+\beta\left(mq_0+(1-m)q_1\right) \Big) \\
&-\frac{\beta q_0}{1-\beta+\beta\big(mq_0+(1-m)q_1\big)}\Big].
\end{split}
\end{equation}
In the second line, the disorder is to be treated as a random walk, with the "steps" $\zeta$ distributed according to $P(\zeta^{\alpha}=x) = \frac{1+\rho}{2}\delta(x-1) + \frac{1-\rho}{2}\delta(x+1)$, and the average computation is detailed in \cref{APP:Average_over_disorder}. As usual, the dominant values of the order parameters are zeroes of the derivatives of $F$ w.r.t. $x_\alpha$, $q_0,q_1, r_0,r_1$ and $m$, and can be found numerically. The saddle point equations read
\begin{equation}
    \begin{aligned}
        &(1-q) \, x_{\alpha} \, + \, q \, \Bigl( \sum_{\alpha'=1}^{K} x_{\alpha'}\Bigr)
        \, (=m^\alpha) = \Big\langle \Big\langle\, \zeta ^\alpha \int {\cal D} z \frac{\int {\cal D} \tilde{z} \cosh^m \tanh}{\int {\cal D} \tilde{z} \cosh^m} \Big\rangle \Big\rangle_{\zeta}\\
        &r_0=\frac{q_0}{\Big(1-\beta+\beta(mq_0+(1-m)q_1)\Big)^2}\\
        &r_1=r_0+\frac{q_1-q_0}{\Big( 1-\beta(1-q_1)\Big)\Big(1-\beta+\beta(mq_0+(1-m)q_1)\Big)}\\
        & q_0 = \Big\langle \Big\langle \int {\cal D} z \Big[ \frac{\int {\cal D} \tilde{z} \cosh^m \tanh}{\int {\cal D} \tilde{z} \cosh^m} \Big]^2 \Big\rangle \Big\rangle_{\zeta} \\
        & q_1 = \Big\langle \Big\langle \int {\cal D} z \frac{\int {\cal D} \tilde{z} \cosh^m \tanh^2}{\int {\cal D} \tilde{z} \cosh^m} \Big\rangle \Big\rangle_{\zeta}\\
        &m= \frac{2}{\alpha\beta} \frac{\Big\langle \Big\langle \int {\cal D} z \frac{\int {\cal D} \tilde{z} \cosh^m \log (\cosh^m) }{\int {\cal D} \tilde{z} \cosh^m} \Big\rangle \Big\rangle_{\zeta} -\Big\langle \Big\langle \int {\cal D} z \log \int {\cal D} \tilde{z} \cosh^m \Big\rangle \Big\rangle_{\zeta}}{\frac{q_0}{1-\beta(1-q_1+m(q_1-q_0))}-\frac{q_1}{1-\beta(1-q_1)}+\frac{1}{\beta m}\log\Big( \frac{1-\beta(1-q_1)}{1-\beta(1-q_1+m(q_1-q_0))}\Big)}
    \end{aligned}
    \label{eq:1RSB_finiteT}
\end{equation}
where we have used the shorthand notation
\begin{equation}
    \begin{aligned}
        &\cosh:=\cosh \Big( \beta\big(\sum_{\alpha=1}^{K} \zeta^{\alpha} x_{\alpha}+ \sqrt{\alpha r_0} \, z + \sqrt{\alpha(r_1-r_0)}\tilde{z}\big)\Big)\\
        &\tanh= \tanh \Big( \beta\big(\sum_{\alpha=1}^{K} \zeta^{\alpha} x_{\alpha} + \sqrt{\alpha r_0} \, z + \sqrt{\alpha(r_1-r_0)}\tilde{z}\big)\Big)
    \end{aligned}
\end{equation}
The corresponding free energy is obtained by evaluating $F[\cdot]$ on the solutions of the saddle-point equations, and the 1RSB expression for the archetype overlaps is
\begin{equation}
O_{\mathrm{1RSB}}
=
\mathbb E_{\boldsymbol\zeta,z}
\left[
    \frac{
        \int\mathcal D\tilde z\,
        \cosh^m(\beta h)\tanh(\beta h)
    }{
        \int\mathcal D\tilde z\,
        \cosh^m(\beta h)
    }
\right]
\end{equation}

%%%%%%%%%%%%%%%%%%%%%%%%%%%%%%%%%%%%%%
\section{Zero Temperature 1RSB free energy}
In the $T\to 0$ limit, as in the Hopfield model, one has $d:=\beta m$ finite, $\delta q_1:=(1-q_1)/T$ finite, and $[\cosh(\beta x)]^m\to \exp(d|x|)$. It follows that, in this limit,
\begin{equation}
\label{eq:T0_1RSB_freeEnergy}
    \begin{aligned}
        F[q_0,\delta q_1, r_0, r_1,d, \, x_{1}, \ldots, x_{K}] &= \frac{1}{2} (1-q) 
        \sum_{\alpha=1}^{K} \bigl(x_{\alpha}\bigr)^{2} 
        + \frac{1}{2} q \, \biggl(\sum_{\alpha=1}^{K} x_{\alpha}\biggr)^{2}
        - \frac{1}{2}\alpha \Big(d(r_0q_0-r_1)-r_1\delta q_1\Big) \\
        &+ \frac{\alpha}{2d}\log\Big(1-\frac{d(1-q_0)}{1-\delta q_1} \Big)- \frac{1}{2}\frac{\alpha q_0}{1-\delta q_1-d(1-q_0)}\\
        & -\frac{1}{d}\Big\langle\Big\langle \int {\cal D} z \log \int {\cal D} \tilde{z} \exp \Big( d \big|\sum_{\alpha=1}^{K} \zeta^{\alpha} x_{\alpha} + \sqrt{\alpha r_0} \, z + b\tilde{z}\, \big|  \Big)  \Big\rangle\Big\rangle_\zeta
    \end{aligned}
\end{equation}
In the last line of \cref{eq:T0_1RSB_freeEnergy}, the innermost integral is a combination of error functions. The corresponding $T\to0$ saddle point equations are obtained by setting to zero the partial derivatives w.r.t. $q_0,\delta q_1, r_0, r_1, d, \, x_{\alpha}$, and can be solved numerically. They read:
\begin{equation}
    \begin{aligned}
        &(1-q) \, x_{\alpha} \, + \, q \, \Bigl( \sum_{\alpha'=1}^{K} x_{\alpha'}\Bigr)= \Big\langle\Big\langle \int {\cal D} z\, \zeta^\alpha\frac{I_2}{I_1}  \Big\rangle\Big\rangle_\zeta\\
        &r_0=\frac{q_0}{(1-\delta q_1-d(1-q_0))^2}\\
        &r_1= r_0+ \frac{1-q_0}{(1-\delta q_1)(1-\delta q_1-d(1-q_0))}\\[0.5em]
        &  q_0=\Big\langle\Big\langle \int {\cal D} z\, \Big(\frac{I_2}{I_1}\Big)^2  \Big\rangle\Big\rangle_\zeta\\[0.5em]
        &\delta q_1=\frac{\sqrt{2}}{\sqrt{\pi}b} \Big\langle\Big\langle \int {\cal D} z\,  \exp\Big( -\frac{A^2}{2}\Big)\frac{1}{I_1}  \Big\rangle\Big\rangle_\zeta\\[0.5em]
        &d= \frac{2}{\alpha} \frac{\Big\langle\Big\langle \int {\cal D} z\,\frac{I_3}{I_1}\Big\rangle\Big\rangle_\zeta - \Big\langle\Big\langle \int {\cal D} z\, \log I_1  \Big\rangle\Big\rangle_\zeta}{\frac{q_0}{1-\delta q_1 - d(1-q_0)}-\frac{1}{1-\delta q_1}+\frac{1}{d}\log\Big( \frac{1-\delta q_1}{1-\delta q_1 - d(1-q_0)}\Big)}
    \end{aligned}
    \label{eq:1RSB_zeroT}
\end{equation}
where
\begin{equation}
\label{eq:useful_identities}
    \begin{aligned}
        &I_1=\int {\cal D} \tilde{z} \exp \big( db|A+\tilde{z}|  \big)\\
        &I_2=\int {\cal D} \tilde{z} \exp \big( db|A+\tilde{z}|  \big) \text{sign} \big( db(A+\tilde{z})  \big)\\
        &I_3= \int {\cal D} \tilde{z} \exp \big( db|A+\tilde{z}|  \big)  db|(A+\tilde{z})|
    \end{aligned}
\end{equation}
and
\begin{equation}
    \begin{aligned}
        &A=\frac{z\sqrt{\alpha r_0}+ \sum_{\alpha'=1}^K \zeta^{\alpha'}x_{\alpha'}}{b}\\
        &b=\sqrt{\alpha(r_1-r_0)}
    \end{aligned}
\end{equation}
The integrals in \cref{eq:useful_identities} result in combinations of error functions.

\section{How to average over disorder}
\label{APP:Average_over_disorder}
A crucial part of the analysis consists in averaging over the quenched disorder. This procedure can be carried out in the same way both within the replica-symmetric approximation and within the one-step replica symmetry breaking scheme. Averaging over the quenched disorder proceeds in two steps. We first perform the average over the variables $\eta$, as they can be integrated out exactly. This step yields an effective description in which the disorder appears only through the variables $\zeta$, as in \cref{eq:RS_F_T!=0,eq:1RSB_F_T!=0}. 
Assuming that $x_2 = x_3 = \cdots = x_K$ as in the main text, we have to compute averages of three kinds: $\langle g(\sum_{\alpha=1}^K\zeta^\alpha x_\alpha)\rangle$; $\langle \zeta^1 g(\sum_{\alpha=1}^K\zeta^\alpha x_\alpha)\rangle$ and $\langle \zeta^2 g(\sum_{\alpha=1}^K\zeta^\alpha x_\alpha)\rangle$, where $g$ is a scalar function arising from the single-site effective measure. It is convenient to rewrite the sum in the following way
\begin{equation}
    \sum_{\alpha=1}^{K} \zeta^{\alpha} x_{\alpha} = \Big(x_1\zeta^1 + x_2 \zeta^2 + x_2 B_K\Big)
\end{equation}
where $B_K = \sum_{\alpha=3}^{K} \zeta^{\alpha}$ collects the contribution of the remaining $K-2$ disorder variables and represents a biased random walk of length $K-2$. By construction, $B_K$ is statistically independent of $\zeta^1$ and $\zeta^2$. With this decomposition, the original high-dimensional quenched average over the $K$ disorder variables $\{\zeta^\alpha\}$ is reduced to a finite sum over the configurations of $(\zeta^1,\zeta^2)$, combined with an average over the collective random-walk variable $B_K$.\\
As an example, let us focus on $\langle g(\sum_{\alpha=1}^K\zeta^\alpha x_\alpha)\rangle$. We first perform the sum over the four possible realizations of $(\zeta^1,\zeta^2)$:
\begin{align}
\label{eq:example_term}
&\left\langle g\left(x_1\zeta^1 + x_2 \zeta^2 + x_2 B_K\right)  \right\rangle = p^2 \left \langle g\left(x_1 + x_2 + x_2 B_K\right) \right \rangle_{B_K} \nonumber\\
&+ p (1-p) (\left \langle g\left(x_1 - x_2 + x_2 B_K\right) \right \rangle_{B_K} + \left \langle g\left(-x_1 + x_2 + x_2 B_K\right) \right \rangle_{B_K})\nonumber\\
&+ (1-p)^2 \left \langle g\left(-x_1 - x_2 + x_2 B_K\right) \right \rangle_{B_K}  \,,
\end{align}
where the symbol $\langle \ldots \rangle_{B_K}$ denotes averaging over realizations of $B_K$ and $p:=P(\zeta=1)=\frac{1+\rho}{2}$.  Averaging over the random walk we have
\begin{equation}
\left \langle g\left(x_1 + x_2 + x_2 B_K\right)  \right \rangle_{B_K} =\frac{(1 - p)^{K - 1}}{p} \sum_{X = - (K-2)}^{K-2} \binom{K-2}{\frac{X + K}{2} - 1} \left(\frac{p}{1 - p}\right)^{(X + K)/2} g\left(x_1 + x_2 + x_2  X\right) 
\end{equation}
where the sum is constrained to $X$ with the same parity of $K$. In the case of $K$ even, $K=2k$ and $X=2n$,
\begin{equation}
\left \langle g\left(x_1 + x_2 + x_2 B_K\right)  \right \rangle_{B_K}=\left(p (1 - p)\right)^{k-1} \sum_{n = - k + 1}^{k-1} \binom{2k-2}{n+k-1} \left(\frac{p}{1 - p}\right)^{n} g\left(x_1 + x_2 + 2 x_2  n\right).
\end{equation}
Analogous expressions hold for the other three terms in \cref{eq:example_term}.
\end{document}